\documentclass[aps,prc,twocolumn, superscriptaddress,showkeys,nofootinbib]{revtex4-1}  

\usepackage[utf8]{inputenc}

\usepackage{graphicx,color}
\usepackage{amsmath,amssymb,amsfonts}
\usepackage{hyperref}

\usepackage{xspace}	

\def\eqref#1{{(\ref{#1})}}

\begin{document}

\title{Constraining the $f_0(980)$ Structure and Formation Dynamics \\
via Anisotropy-Response Scaling}

\author{Roy~Lacey} 
\affiliation{Department of Chemistry, the State University of New York, Stony Brook, New York 11794, USA}
\affiliation{Department of Physics, Stony Brook University, Stony Brook, NY 11794}

\begin{abstract}
A species-resolved anisotropy-response scaling framework is used to
investigate the structure and formation dynamics of the $f_0(980)$ in
relativistic nuclear collisions. Published measurements in
high-multiplicity p+Pb collisions at $\sqrt{s_{NN}}=8.16$~TeV exhibit
broad scaling closure under a single-meson response construction, with
a small effective final-state response, $\zeta_{f_0}\simeq0.04$, whereas
a symmetric ${\rm K\bar K}$ constituent-response construction gives
substantially poorer closure. An explicitly molecular ${\rm K\bar K}$
calculation also exhibits single-meson-like scaling but gives a
substantially larger effective final-state response,
$\zeta_{f_0}\simeq0.16$. Thus, single-meson scaling alone is not a
unique structural discriminator. The combined response construction
and $\zeta_{f_0}$ constraints characterize the measured $f_0(980)$ by
a single-meson-like anisotropy response with weak final-state
sensitivity, distinctly different from the molecular benchmark.
These results establish anisotropy-response scaling as a new
experimental probe of $f_0(980)$ formation dynamics and motivate
systematic tests across collision systems and beam energies.
\end{abstract}

\maketitle

\section{Introduction}
\label{sec:introduction}

The internal structure of the $f_0(980)$ remains a long-standing
question in hadron spectroscopy. Its mass lies remarkably close to the
${\rm K\bar K}$ threshold, and its properties have motivated
interpretations ranging from a conventional $q\bar q$ meson to a
compact tetraquark state and a predominantly ${\rm K\bar K}$ hadronic
molecule
\cite{Jaffe:1976ig,Weinstein:1982gc,Weinstein:1990gu,Oller:1997ti,Amsler:2004ps}.
The near-threshold location is especially significant for the
molecular interpretation, for which the relevant degrees of freedom
are two correlated kaons rather than those of a single compact meson.
These alternatives therefore imply potentially different formation
histories and medium responses, even when they lead to the same
observed resonance. Observables that retain sensitivity to how the
$f_0(980)$ acquires its momentum anisotropy can consequently provide
structural information complementary to that obtained from its mass,
decay widths, and branching fractions alone.

Relativistic nuclear collisions provide such an opportunity because
the azimuthal anisotropy of identified hadrons reflects both their
coupling to the evolving medium and the dynamics governing their
formation. The approximate number-of-constituent-quark (NCQ) scaling
of elliptic flow, $v_2$, observed for identified hadrons has long been
interpreted as evidence that a substantial fraction of the anisotropy
is established at the partonic stage and subsequently inherited
through hadronization
\cite{Molnar:2003ff,Fries:2003vb,Greco:2003mm,STAR:2004jwm,PHENIX:2006dpn}.
More recently, the CMS Collaboration measured the $v_2(p_T)$ of the
$f_0(980)$ in high-multiplicity p+Pb collisions at
$\sqrt{s_{NN}}=8.16$~TeV and compared its NCQ-scaled behavior with
that of established mesons and baryons \cite{CMS:2023rev}. The
comparison favors a two-constituent assignment over four-quark and
three-constituent alternatives, providing evidence for meson-like
behavior of the $f_0(980)$ in these collisions.

The structural interpretation of this observation is not unique.
Recent calculations show that the measured $f_0(980)$ elliptic flow
can also be described within a ${\rm K\bar K}$ molecular picture
\cite{Wang:2025sxa}. In this approach, the $f_0(980)$ is formed
through the coalescence of kaons near kinetic freeze-out, so that its
final momentum and anisotropy are inherited from the phase-space
distributions of the two constituent kaons. Importantly, realistic
coalescence need not correspond to equal momentum sharing between the
kaons; consequently, the simple constituent expectation
$v_2^{f_0}(P_T)\simeq2v_2^K(P_T/2)$ need not hold even for a
predominantly molecular state. Thus, neither meson-like NCQ scaling
nor the validity of a simple two-kaon constituent mapping provides,
by itself, a unique determination of the microscopic structure. A
more restrictive test requires consideration of the full
species-dependent response.

The present study addresses this question using a species-resolved
anisotropy-response scaling framework developed to separate geometric,
attenuation, radial-flow, and late-stage hadronic contributions to the
measured anisotropy
\cite{Lacey:2024uky,Lacey:2024bcm,Lacey:2026gqe}. A central feature
of the approach is that scaling is not treated as a fit: the common
response parameters are constrained through the simultaneous scaling
of established identified-hadron species before the $f_0(980)$
response is examined. Three complementary diagnostics are then
employed. The $f_0(980)$ is tested under a single-meson response
construction using its physical mass and measured transverse momentum,
and under a symmetric ${\rm K\bar K}$ constituent-response limit using
equal constituent momentum sharing and the kaon mass scale. After the
common system response is constrained, the effective $f_0(980)$
final-state response required for scaling closure is determined.
Together, these diagnostics constrain whether the observed $f_0(980)$
anisotropy follows a single-meson or symmetric two-kaon constituent
response, as well as its sensitivity to final-state dynamics.

The microscopic ${\rm K\bar K}$ coalescence calculation of
Ref.~\cite{Wang:2025sxa} provides an important benchmark because its
molecular structure is specified independently of the response
scaling. Applying the same diagnostics to the calculated kaon and
$f_0(980)$ anisotropies tests how an explicitly molecular state is
represented within the response framework. In particular, it
determines whether molecular formation necessarily produces symmetric
two-kaon constituent-response scaling and establishes the effective
$f_0(980)$ final-state response required after the common system
response is constrained by the calculated kaons.

The same diagnostics are applied to the published $f_0(980)$
$v_2(p_T)$ measurements in high-multiplicity p+Pb collisions at
$\sqrt{s_{NN}}=8.16$~TeV \cite{CMS:2023rev}. The broad measured
$p_T$ range provides sensitivity to the evolution from collective
dynamics at low $p_T$ toward a regime in which parton energy loss
and jet-related contributions become increasingly important
at higher $p_T$. Because the system response is independently
constrained by established identified hadrons, the response
construction and effective $f_0(980)$ final-state response provide
complementary information beyond constituent counting alone.

\section{Anisotropy-Response Scaling Framework}
\label{sec:framework}
The analysis employs the established species-resolved
anisotropy-response scaling framework
\cite{Lacey:2024uky,Lacey:2024bcm,Lacey:2026gqe}, in which the
measured harmonic coefficients are expressed as reduced responses,
$v_n(p_T)/\varepsilon_n$, and mapped onto a common scaling function
after scaling out contributions associated with the initial geometry,
system size, viscous attenuation, radial flow, and late-stage hadronic
re-scattering. The species-resolved scaling is anchored to the kaon
scaling function in ultra-central (0--1\%) Pb+Pb collisions at
$\sqrt{s_{NN}}=5.02$~TeV, which provides the common meson reference.
The characteristic transverse size is represented by
$\mathcal{R}\propto\langle N_{\rm chg}\rangle^{1/3}$. The attenuation
scale $\beta_0$ is fixed by the ultra-central Pb+Pb reference, while
the attenuation strength for a comparison system is expressed as
$\beta=k_\beta\beta_0$. The momentum-dependent viscous correction is
represented by $\delta f=\kappa p_T^2$. Hadronic re-scattering and 
radial-flow contributions are encoded by
$\zeta_{\rm hs}$ and $\zeta_{\rm rf}$, respectively, with the
corresponding meson and baryon response factors
$\zeta_m=1-\zeta_{\rm hs}$ and $\zeta_b=(1-\zeta_{\rm rf})^{|n_B|}$.

The parameters governing these transformations are constrained through
the simultaneous scaling of established identified-hadron species.
The resulting common response provides the baseline against which the
$f_0(980)$ is tested. Structural information is introduced through
the response construction applied to the measured $f_0(980)$ rather
than through independent modification of the common system-response
parameters. The analysis considers a single-meson response and a
symmetric ${\rm K\bar K}$ constituent-response limit. After the common 
system response has been constrained, scaling closure
determines the effective $f_0(980)$ final-state response. The response
construction and effective final-state response therefore provide
complementary diagnostics of $f_0(980)$ formation dynamics.

\subsection{Common meson response}
\label{sec:meson_response}
For identified mesons, the intra-harmonic $2\!\to\!2$ relation maps
the reduced elliptic response of a comparison system onto the
ultra-central kaon reference according to \cite{Lacey:2026gqe}
\begin{multline}
\left(\frac{v_2(p_T)}{\varepsilon_2}\right)_{\rm uc}
e^{\frac{2\beta_0}{\mathcal{R}_{\rm uc}}(2+\kappa p_T^2)}
=
e^{\frac{\beta_0}{\mathcal{R}_{\rm uc}}
A_\alpha(p_T)\zeta_M^{(2)}}\times
\\
\left(\frac{v_2'(p_T)}{\varepsilon_2'}\right)^{\zeta_m}
e^{\frac{2\zeta_m\beta}{\mathcal{R}_{\rm uc}}
\left(\frac{\mathcal{R}_{\rm uc}}{\mathcal{R}'}-1\right)
(2+\kappa p_T^2)} .
\label{eq:meson_v2_scaling}
\end{multline}
where the left-hand side defines the ultra-central kaon reference
response and the right-hand side transforms the measured meson response
from the comparison system and centrality onto that reference. Primed 
quantities denote the comparison system, with $\mathcal{R}'$
and $\varepsilon_2'$ specifying its characteristic size and initial
ellipticity, respectively. Scaling closure is obtained when the
transformed response collapses onto the common reference response.

The meson normalization factor is
$\zeta_M^{(2)}=\zeta_m+\gamma_{32}^{X}+(1-k_\beta)$. The
geometry-only offset
$\gamma_{32}^{X}\equiv
\ln[(\varepsilon_3/\varepsilon_2)_{\rm ref}/
(\varepsilon_3/\varepsilon_2)_{\rm sys}]|_X$
is inherited from the simultaneous intra- and inter-harmonic scaling
framework and is determined from the corresponding eccentricity ratios
rather than varied as an independent medium-response parameter. For
p+Pb collisions, the centrality evolution of
$\varepsilon_3/\varepsilon_2$ is inverted relative to that of the
large-$A+A$ reference systems. Accordingly, the sign of
$\gamma_{32}^{X}$ is reversed for the p+Pb scaling to preserve the
orientation of the geometry correction. The remaining response
parameters are unchanged.

The normalization appropriate to the different centrality classes is
specified by
\begin{equation}
A_\alpha(p_T)=
\begin{cases}
2, & \alpha=0,\\[1mm]
2\alpha(2+\kappa p_T^2), & \alpha>0,
\end{cases}
\label{eq:Aalpha}
\end{equation}
where $\alpha=0$, $0.5$, and $1$ define the non-ultra-central
(non-UC), near-ultra-central (near-UC), and ultra-central (UC)
scaling selections, respectively. The attenuation term proportional
to $\beta/\mathcal{R}$ retains its full $(2+\kappa p_T^2)$ dependence
for all centrality selections.

For cross-species comparisons, the transverse kinetic energy
${\rm KE}_T=m_T-m_0$, with $m_T=\sqrt{p_T^2+m_0^2}$, is used in place
of $p_T$, following the species-resolved prescription
\cite{Lacey:2026gqe}. This representation scales out the leading
kinematic mass dependence and provides the common variable for
comparing identified-hadron responses.

Equation~(\ref{eq:meson_v2_scaling}) therefore provides the common
meson-response kernel for the present analysis, anchored to the
ultra-central kaon reference. The same response parameters are used
when the $f_0(980)$ is subsequently examined under the single-meson
and symmetric ${\rm K\bar K}$ constituent-response constructions.

\subsection{Baryon response}
\label{sec:baryon_response}
Identified baryons provide additional controls on the
species-resolved scaling and independently constrain the common
response. Their anisotropy is mapped onto the same ultra-central kaon
reference using the common geometry, system-size, attenuation, and
viscous-response parameters, while the species-dependent radial-flow
response is encoded through
$\zeta_b=(1-\zeta_{\rm rf})^{|n_B|}$, where $n_B$ is the baryon
number.

The corresponding intra-harmonic $2\!\to\!2$ baryon relation is
\begin{multline}
\left(\frac{v_2(p_T)}{\varepsilon_2}\right)_{\rm uc}
e^{\frac{2\beta_0}{\mathcal{R}_{\rm uc}}(2+\kappa p_T^2)}
=
e^{(1-\alpha)\frac{2\beta_0}{\mathcal{R}_{\rm uc}}
(2+\kappa p_T^2)\zeta_B^{(2)}}\times
\\
\left(\frac{v_2'(p_T)}{\varepsilon_2'}\right)^{\zeta_b}
e^{\frac{2\zeta_b\beta}{\mathcal{R}_{\rm uc}}
\left(\frac{\mathcal{R}_{\rm uc}}{\mathcal{R}'}-1\right)
(2+\kappa p_T^2)},
\label{eq:baryon_v2_scaling}
\end{multline}
where
$\zeta_B^{(2)}=-\zeta_b(|n_B|/k_\beta-\gamma_{32}^{X})$.
The parameter $\alpha$ retains the same centrality assignments defined
for the meson scaling but enters the baryon normalization through the
factor $(1-\alpha)$.

As in the meson relation, the left-hand side defines the common
ultra-central kaon reference response, while the right-hand side maps
the measured baryon response from the comparison system onto that
reference. The baryon-specific factor $\zeta_b$ scales out the
radial-flow response and its baryon-number dependence; the remaining
geometry, attenuation, and system-size inputs are common to the meson
and baryon sectors.

For cross-species comparisons, the same
${\rm KE}_T=m_T-m_0$ representation is used. The baryon measurements
therefore provide a complementary constraint on the common response
and, in particular, on $\zeta_{\rm rf}$. Together with the established
meson scaling, they define the system-response baseline against which
the $f_0(980)$ response is subsequently assessed.

\subsection{\texorpdfstring{$f_0(980)$}{f0(980)} response constructions}
\label{sec:f0_response}
With the common system response constrained by the established meson
and baryon scaling, the $f_0(980)$ can be examined without redefining
the underlying geometry, system-size, attenuation, or viscous-response
parameters. The analysis considers two distinct constructions of the
measured $f_0(980)$ anisotropy: a single-meson response and a symmetric
${\rm K\bar K}$ constituent-response limit. These constructions test whether 
the measured $f_0(980)$ anisotropy follows a single-meson or symmetric 
two-kaon constituent response within the same scaling framework. 
Scaling closure then determines the effective $f_0(980)$ final-state
response within the established species-resolved framework.

\subsubsection{Single-meson response}
\label{sec:single_f0}
In the single-meson construction, the measured $f_0(980)$ is evaluated
with its physical transverse momentum and mass, $p_T=P_T$ and
$m_0=m_{f_0}$, so that
${\rm KE}_T^{f_0}=\sqrt{P_T^2+m_{f_0}^2}-m_{f_0}$. The resulting
reduced response is transformed with the common meson kernel of
Eq.~(\ref{eq:meson_v2_scaling}). This construction tests whether the
measured $f_0(980)$ anisotropy follows the common single-meson response
after the system-dependent contributions have been scaled out.

Agreement with the single-meson response does not, by itself, uniquely
determine the microscopic $f_0(980)$ wave function. In particular, a
state produced through more complex late-stage formation dynamics can
exhibit the same response if its measured momentum and anisotropy
satisfy the single-meson scaling relation.

\subsubsection{Symmetric \texorpdfstring{${\rm K\bar K}$}{K-Kbar} constituent-response limit}
\label{sec:kk_limit}
A two-kaon composite provides a distinct limiting response. At small
anisotropy, the elliptic anisotropy of a composite formed from a kaon
and an antikaon can be expressed schematically as
\begin{equation}
v_2^{f_0}(P_T)
\simeq
v_2^K(p_{T1})+v_2^{\bar K}(p_{T2}),
\qquad
p_{T1}+p_{T2}=P_T .
\label{eq:kk_general}
\end{equation}
For symmetric constituent momentum sharing,
$p_{T1}=p_{T2}=P_T/2$, and approximately equal kaon and antikaon
responses, this reduces to
\begin{equation}
\frac{v_2^{f_0}(P_T)}{2}
\simeq
v_2^K(P_T/2).
\label{eq:kk_symmetric}
\end{equation}
The corresponding response mapping is
$P_T\rightarrow P_T/2$,
$v_2^{f_0}\rightarrow v_2^{f_0}/2$, and $m_0\rightarrow m_K$.
Consequently, the constituent transverse kinetic energy used in the
cross-species comparison is
${\rm KE}_T^K=\sqrt{(P_T/2)^2+m_K^2}-m_K$.

The symmetric ${\rm K\bar K}$ mapping defines an equal-momentum
constituent-response limit rather than a general scaling requirement
for a molecular state. More generally, molecular formation can involve
unequal momentum sharing between the two kaons, for which the measured
$f_0(980)$ anisotropy need not reduce to
Eq.~(\ref{eq:kk_symmetric}). The symmetric construction therefore
provides a well-defined limiting test against the common kaon response
without requiring all ${\rm K\bar K}$ molecular formation to satisfy
this mapping.

\subsubsection{Effective \texorpdfstring{$f_0(980)$}{f0(980)} final-state response}
\label{sec:f0_late_response}
The established species-resolved framework incorporates
species-dependent final-state response through the same scaling
parameters used to obtain closure for identified hadrons. For mesons,
$\zeta_{\rm hs}$ is constrained principally through species such as
pions that are sensitive to late-stage hadronic re-scattering, with
$\zeta_m=1-\zeta_{\rm hs}$. For baryons, the characteristic radial-flow
blue shift is scaled out through $\zeta_{\rm rf}$, entering through
$\zeta_b=(1-\zeta_{\rm rf})^{|n_B|}$. These quantities therefore
constrain distinct components of the species-dependent final-state
response.

For the $f_0(980)$, the relative sensitivity to these effects cannot
be assigned a priori because its formation history and interaction
with the evolving medium are themselves part of the physics under
investigation. The effective $f_0(980)$ final-state response is
therefore denoted by $\zeta_{f_0}$ and is determined only after the
common geometry, system-size, attenuation, and viscous-response
contributions have been constrained by the established identified
hadrons.

The quantity $\zeta_{f_0}$ does not represent a new attenuation
mechanism or an independent modification of the system-wide response.
Rather, it quantifies the effective $f_0(980)$ sensitivity to the
final-state dynamics represented by $\zeta_{\rm hs}$ and
$\zeta_{\rm rf}$, including possible contributions from hadronic
re-scattering, radial flow, formation, and regeneration. No particular
decomposition of $\zeta_{f_0}$ into these contributions is assumed.

The $f_0(980)$ is therefore constrained by two complementary
diagnostics: the scaling fidelity of the single-meson and symmetric
${\rm K\bar K}$ constituent-response constructions and the magnitude
of $\zeta_{f_0}$ obtained from scaling closure.

\subsection{Molecular-response benchmark}
\label{sec:molecular_benchmark}
The ${\rm K\bar K}$ coalescence calculation of
Ref.~\cite{Wang:2025sxa} provides a molecular-response benchmark for
interpreting the $f_0(980)$ response constructions. In this calculation,
the $f_0(980)$ is specified independently as a ${\rm K\bar K}$
molecular state formed from kaons near kinetic freeze-out. The kaon
phase-space distributions are generated within a framework combining
hydrodynamic evolution, coalescence, fragmentation, and hadronic
transport, after which the molecular $f_0(980)$ is constructed from
the resulting kaon distributions. The calculated kaon and $f_0(980)$
elliptic anisotropies in 0--20\% p+Pb collisions at
$\sqrt{s_{NN}}=5.02$~TeV therefore provide a benchmark in which the
molecular character is specified independently of the present response
analysis. The calculation is used as a molecular-response benchmark
rather than as a collision-condition-matched description of the CMS
measurement; in each case, the common system response is constrained
independently before the $f_0(980)$ response is examined.

This benchmark is particularly useful because realistic molecular
coalescence does not require symmetric constituent momentum sharing.
For a finite molecular wave function, kaons with unequal transverse
momenta can contribute to an $f_0(980)$ of momentum $P_T$.
Ref.~\cite{Wang:2025sxa} demonstrates that, for molecular radii in the
physically relevant range, this unequal momentum sharing leads to
violations of the simple symmetric constituent scaling of
Eq.~(\ref{eq:kk_symmetric}). The symmetric limit is approached only
for sufficiently large molecular radii and correspondingly narrow
relative-momentum distributions. Thus, departure from the symmetric
${\rm K\bar K}$ mapping does not, by itself, exclude a molecular
formation mechanism.

The molecular ${\rm K\bar K}$ calculation of
Ref.~\cite{Wang:2025sxa} is subjected to the same response hierarchy
used for the experimental analysis. The calculated kaon anisotropy
first constrains the system-wide attenuation response through
$\beta=k_\beta\beta_0$, which is then held fixed when the molecular
$f_0(980)$ is examined. The $f_0(980)$ is tested under both the
single-meson and symmetric ${\rm K\bar K}$ constituent-response
constructions, while its effective final-state response is quantified
through $\zeta_{f_0}$.

\begin{figure*}[t]
\centering
\includegraphics[width=0.75\textwidth]
{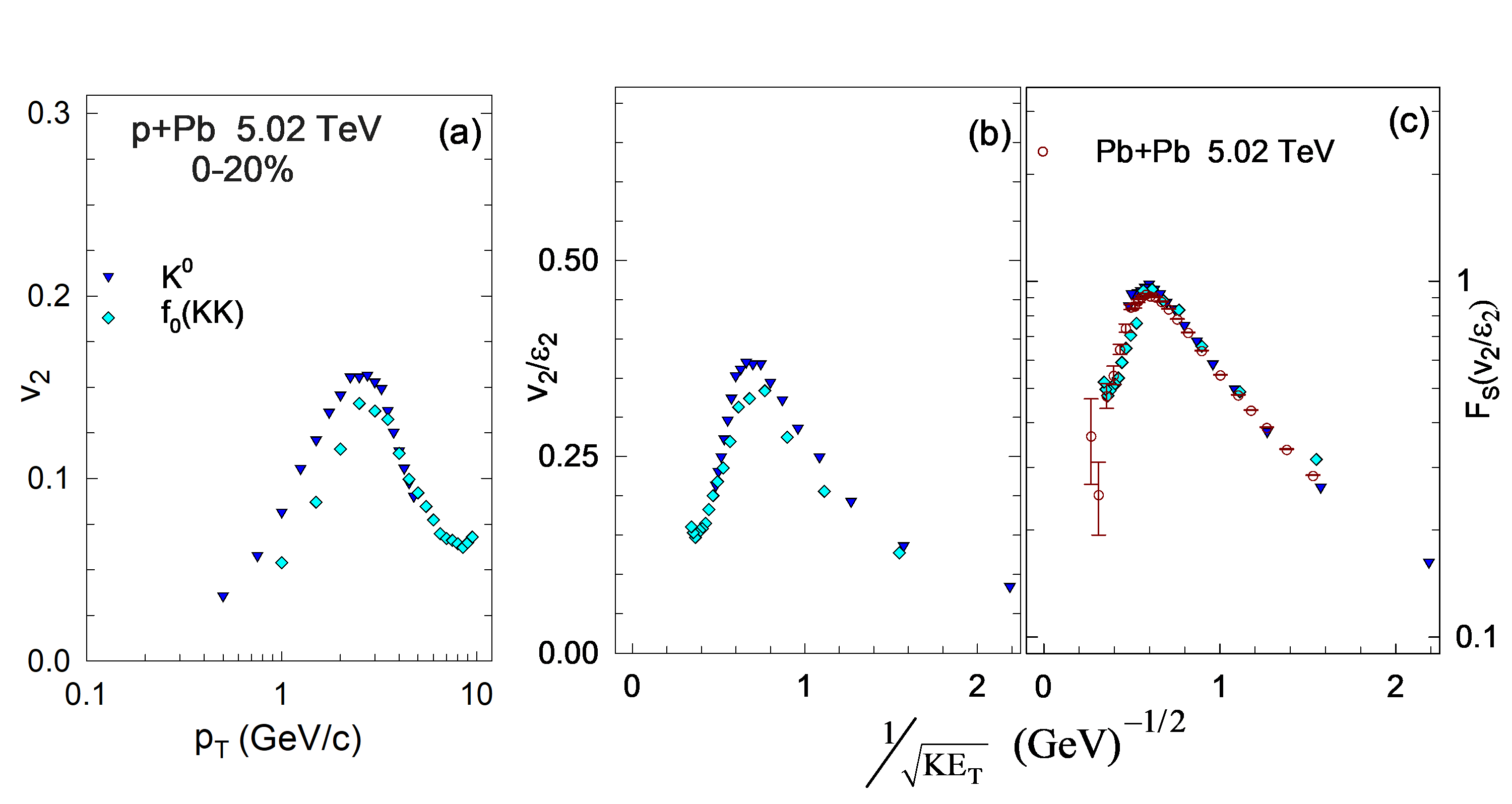}
\includegraphics[width=0.75\textwidth]
{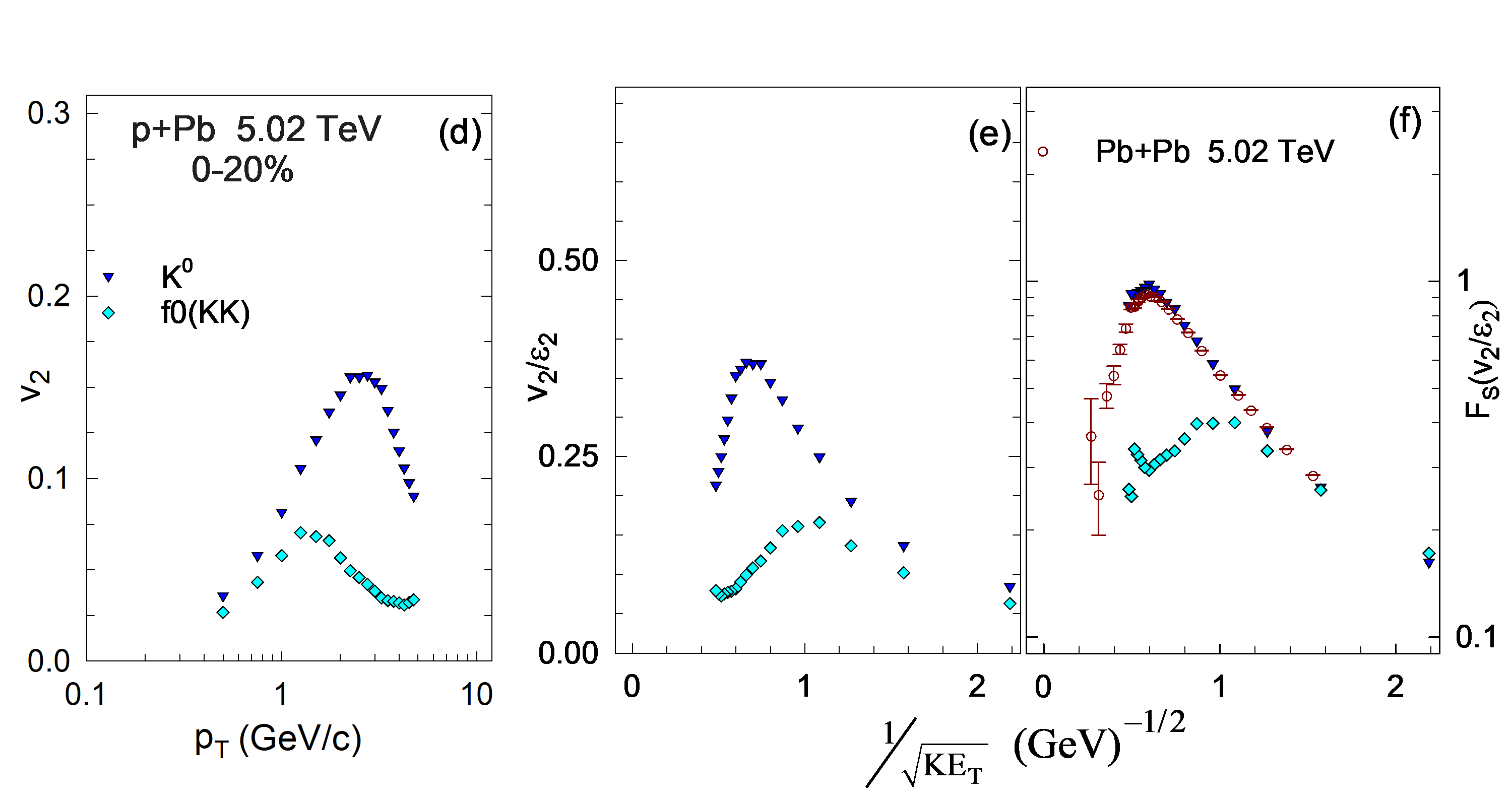}
\caption{
Anisotropy-response scaling of the kaon and molecular $f_0(980)$
calculations of Ref.~\cite{Wang:2025sxa} for 0--20\% p+Pb collisions
at $\sqrt{s_{NN}}=5.02$~TeV. Panels (a)--(c) show the single-meson
$f_0(980)$ construction and panels (d)--(f) the symmetric
${\rm K\bar K}$ constituent-response construction. From left to right,
the panels show $v_2(p_T)$, $v_2/\varepsilon_2$, and the fully
transformed scaling functions; panels (b), (c), (e), and (f) are shown
as functions of $1/\sqrt{{\rm KE}_T}$. The open circles denote the
ultra-central Pb+Pb reference.
}
\label{fig:wang_scaling}
\end{figure*}

Figure~\ref{fig:wang_scaling} shows the resulting response tests.
Panels (a)--(c) treat the molecular $f_0(980)$ under the single-meson
construction, whereas panels (d)--(f) apply the symmetric
${\rm K\bar K}$ constituent-response construction. Despite its
explicitly molecular origin, the calculated $f_0(980)$ exhibits good
scaling closure over a broad $1/\sqrt{{\rm KE}_T}$ range under the
single-meson construction. The symmetric ${\rm K\bar K}$ construction
shows approximate agreement with the common response at low
${\rm KE}_T$, but develops a pronounced separation with increasing
${\rm KE}_T$ that persists after the full response transformation.

For this calculation, the kaon response constrains the common
attenuation strength to $k_\beta\simeq0.55$ and exhibits essentially
no effective hadronic re-scattering response. With this system response
held fixed, scaling closure of the molecular $f_0(980)$ under the
single-meson construction gives a substantial effective final-state
response, $\zeta_{f_0}\simeq0.16$. Thus, single-meson-like anisotropy
scaling does not uniquely identify a compact microscopic state, and
departure from the symmetric ${\rm K\bar K}$ constituent-response
limit does not, by itself, exclude a molecular state. Importantly,
$\zeta_{f_0}\simeq0.16$ provides a quantitative response benchmark
for an explicitly molecular $f_0(980)$ after the common system
response has been independently constrained by the calculated kaons.

\section{Data and Analysis}
\label{sec:data_analysis}
The primary $f_0(980)$ input is the published CMS measurement of
$v_2(p_T)$ in high-multiplicity p+Pb collisions at
$\sqrt{s_{NN}}=8.16$~TeV \cite{CMS:2023rev}. The broad measured
transverse-momentum range permits the $f_0(980)$ response to be tested
from the low-$p_T$ collective region toward higher momentum, where
parton energy loss and jet-related contributions become increasingly
important. The measurement is examined under the single-meson and
symmetric ${\rm K\bar K}$ constituent-response constructions defined
in Sec.~\ref{sec:f0_response}.

\subsection{Data selection and scaling inputs}
\label{sec:data_inputs}
The p+Pb analysis uses identified-hadron $v_2(p_T)$ measurements for
$K_S^0$, $D^0$, $f_0(980)$, $\Lambda$, $\Xi$, and $\Omega$ in the
high-multiplicity event class. The $K_S^0$ measurement provides the
principal light-meson reference, while the identified baryons
constrain the baryon response and its radial-flow dependence. The
$D^0$ meson provides an additional meson-sector comparison at
substantially larger mass. The light-meson and baryon measurements
therefore establish the collision-system response before the
$f_0(980)$ is examined.

The initial-state eccentricities $\varepsilon_n$, charged-particle
multiplicities $\langle N_{\rm chg}\rangle$, and corresponding
system-size inputs are taken from the established species-resolved
scaling analyses
\cite{Lacey:2024uky,Lacey:2024bcm,Lacey:2025bos}.

For each species, the measured $v_2(p_T)$ is reduced by the
corresponding initial-state eccentricity $\varepsilon_2$ and
transformed according to the meson or baryon scaling prescriptions of
Eqs.~(\ref{eq:meson_v2_scaling}) and
(\ref{eq:baryon_v2_scaling}). The characteristic size
$\mathcal{R}$ is determined from the charged-particle multiplicity,
while the attenuation coefficient $k_\beta$ and the response
parameters $\zeta_{\rm hs}$ and $\zeta_{\rm rf}$ are constrained by
the established identified-hadron scaling. The viscous parameter
$\kappa$ retains the value used in the established species-resolved
analysis. These inputs and common response parameters are not
independently readjusted for the $f_0(980)$.
\begin{figure*}[t]
\centering
\includegraphics[width=0.75\textwidth]
{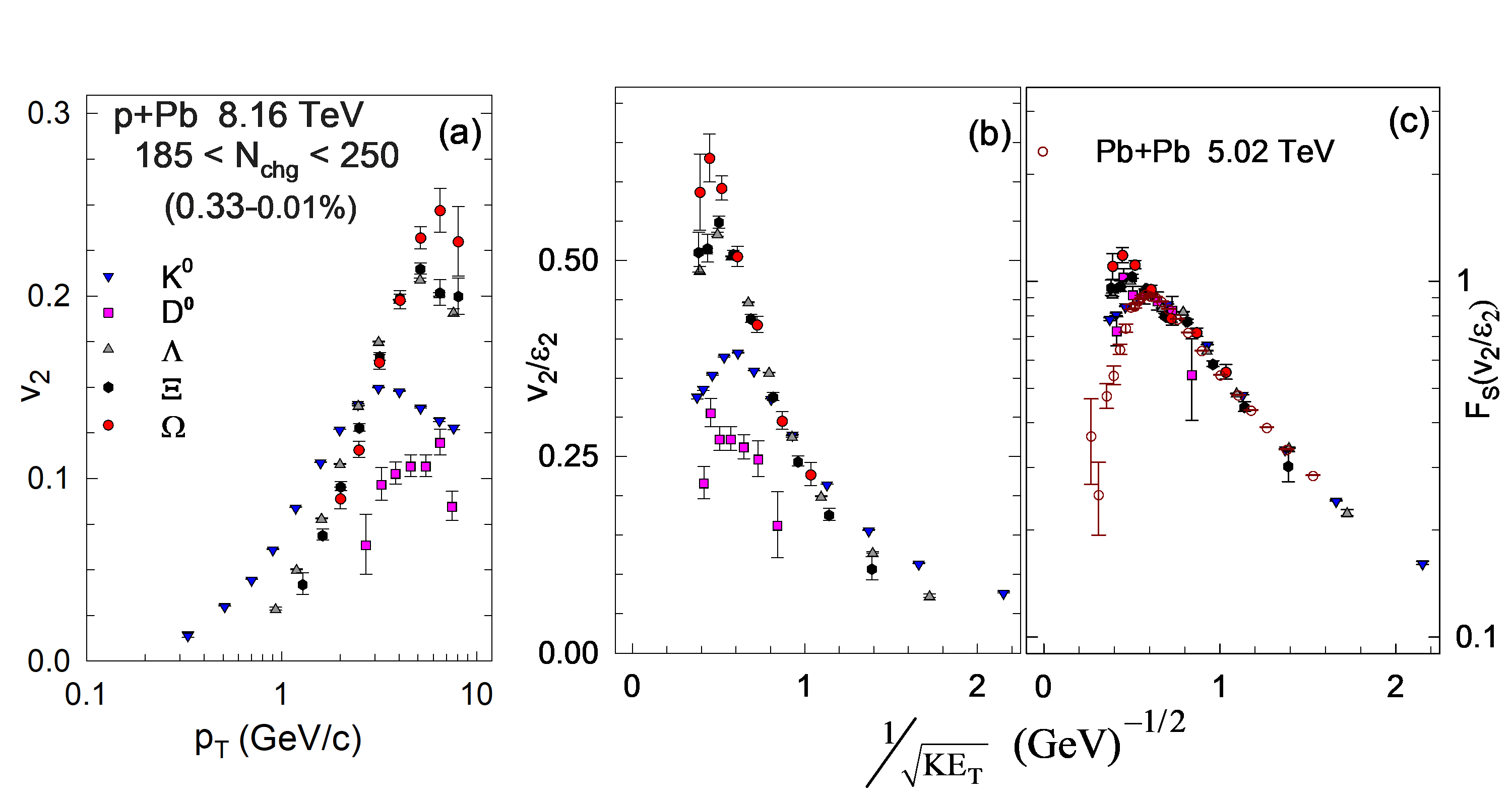}
\caption{Species-resolved anisotropy scaling for $K_S^0$, $D^0$,
$\Lambda$, $\Xi$, and $\Omega$ in high-multiplicity p+Pb collisions
at $\sqrt{s_{NN}}=8.16$~TeV. Shown are (a) $v_2(p_T)$,
(b) the reduced responses $v_2/\varepsilon_2$, and (c) the fully
transformed responses compared with the ultra-central Pb+Pb kaon
reference.}
\label{fig:pid_scaling_ppb}
\end{figure*}

Cross-species comparisons use ${\rm KE}_T=m_T-m_0$. In the
single-meson construction, the measured $f_0(980)$ is evaluated with
$p_T=P_T$ and $m_0=m_{f_0}$. In the symmetric ${\rm K\bar K}$
constituent-response construction, the same measured points are mapped
according to $P_T\rightarrow P_T/2$,
$v_2^{f_0}\rightarrow v_2^{f_0}/2$, and $m_0\rightarrow m_K$.
Thus, the two tests differ only in the response construction applied
to the same measured $f_0(980)$ anisotropy.

\subsection{Momentum-domain treatment}
\label{sec:momentum_domain}
The momentum range used in the scaling is separated according to the
response behavior established for the identified-hadron species. In
large collision systems, the transition from predominantly collective
behavior toward a region with increasing sensitivity to parton energy
loss occurs near $p_T\sim4.5$~GeV/$c$. In p+Pb collisions, the
corresponding transition occurs at substantially lower momentum,
reflecting the earlier departure from the predominantly collective
response in the smaller collision system. The smaller system size also
implies substantially reduced partonic energy loss and hence a weaker
quenching contribution to the anisotropy at higher $p_T$, while the
reduced jet suppression increases the relative importance of
jet-related correlations and nonflow.

For the p+Pb scaling, characteristic transition scales of
$p_T\simeq1.6$~GeV/$c$ for mesons and
$p_T\simeq3.0$~GeV/$c$ for baryons are used. These scales are
constrained by the observed departures from the predominantly
collective response for the respective species classes in p+Pb
collisions \cite{Lacey:2024uky,Lacey:2024bcm,Lacey:2025bos}, rather
than adjusted for the present $f_0(980)$ analysis. They therefore
identify the momentum at which the predominantly collective-response
prescription is replaced by the high-$p_T$ response prescription,
rather than the onset of a quenching-dominated regime. The meson
prescription is applied consistently to the light-meson response and
to the single-meson $f_0(980)$ construction, while the baryon
prescription is applied to $\Lambda$, $\Xi$, and $\Omega$.

The viscous contribution follows the established prescription,
$\delta f=\kappa p_T^2$ in the collective-flow region and is held
fixed at its transition value in the high-$p_T$ response region.
For the symmetric ${\rm K\bar K}$ construction, the meson transition
scale is applied after the constituent-momentum mapping defined above.
The transition scales do not otherwise modify the scaling
prescription.

\subsection{Scaling implementation}
\label{sec:scaling_implementation}
The scaling analysis is performed sequentially. The established
identified-hadron species are first used to constrain the common
system response through the meson and baryon scaling relations of
Sec.~\ref{sec:framework}. These constraints, together with the
geometry and system-size inputs and the established viscous-response
parameter $\kappa$, define the common response used for the subsequent
$f_0(980)$ analysis.

The measured $f_0(980)$ points are then subjected to the single-meson
and symmetric ${\rm K\bar K}$ response constructions without
readjusting the common system-response parameters. Scaling closure 
then determines the effective $f_0(980)$ final-state response $\zeta_{f_0}$, 
as defined in Sec.~\ref{sec:f0_late_response}.

The resulting scaling fidelity and $\zeta_{f_0}$ are then compared
with the explicitly molecular benchmark of
Sec.~\ref{sec:molecular_benchmark}.

\begin{figure*}[t]
\centering
\includegraphics[width=0.75\textwidth]
{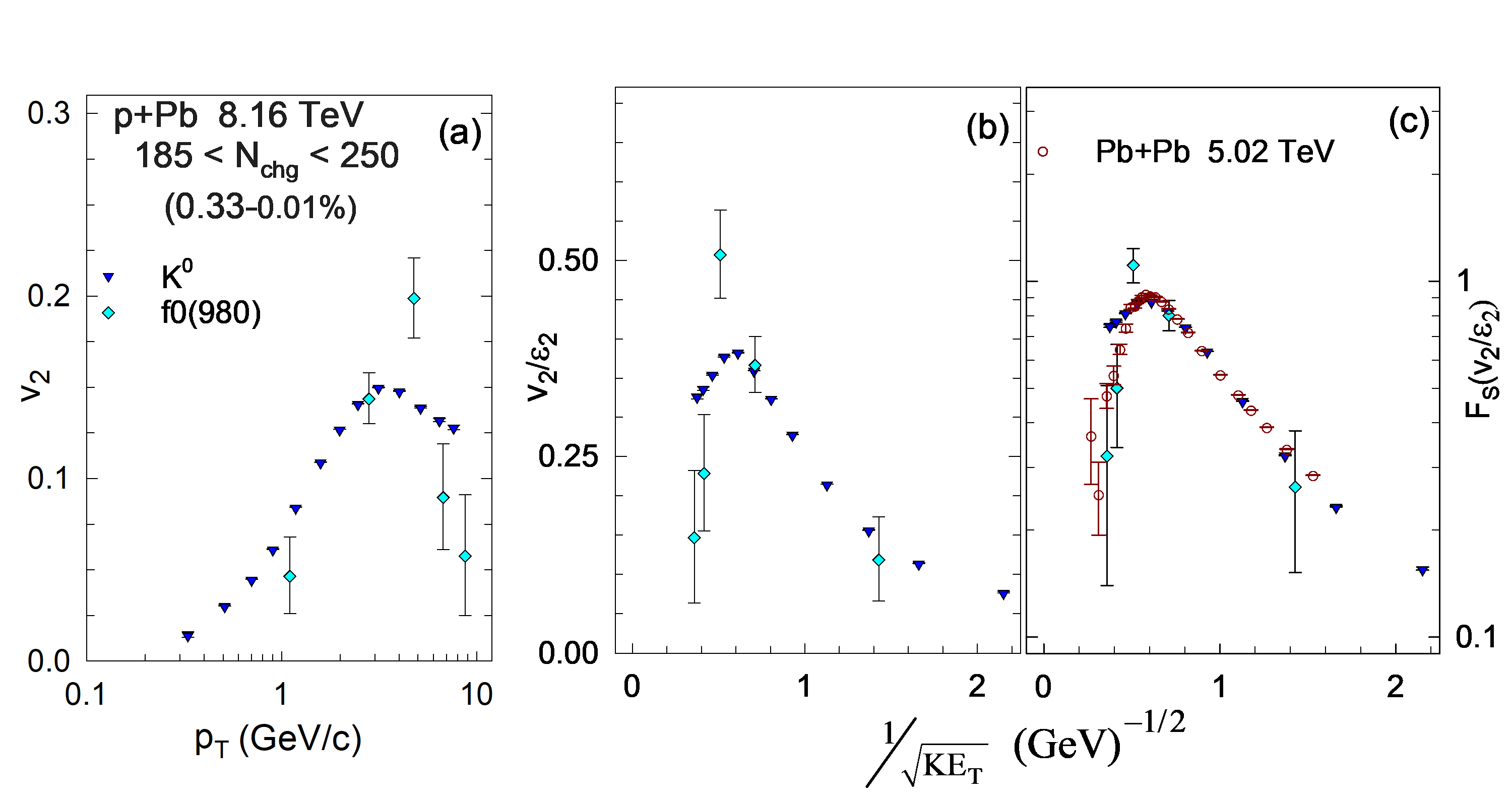}
\includegraphics[width=0.75\textwidth]
{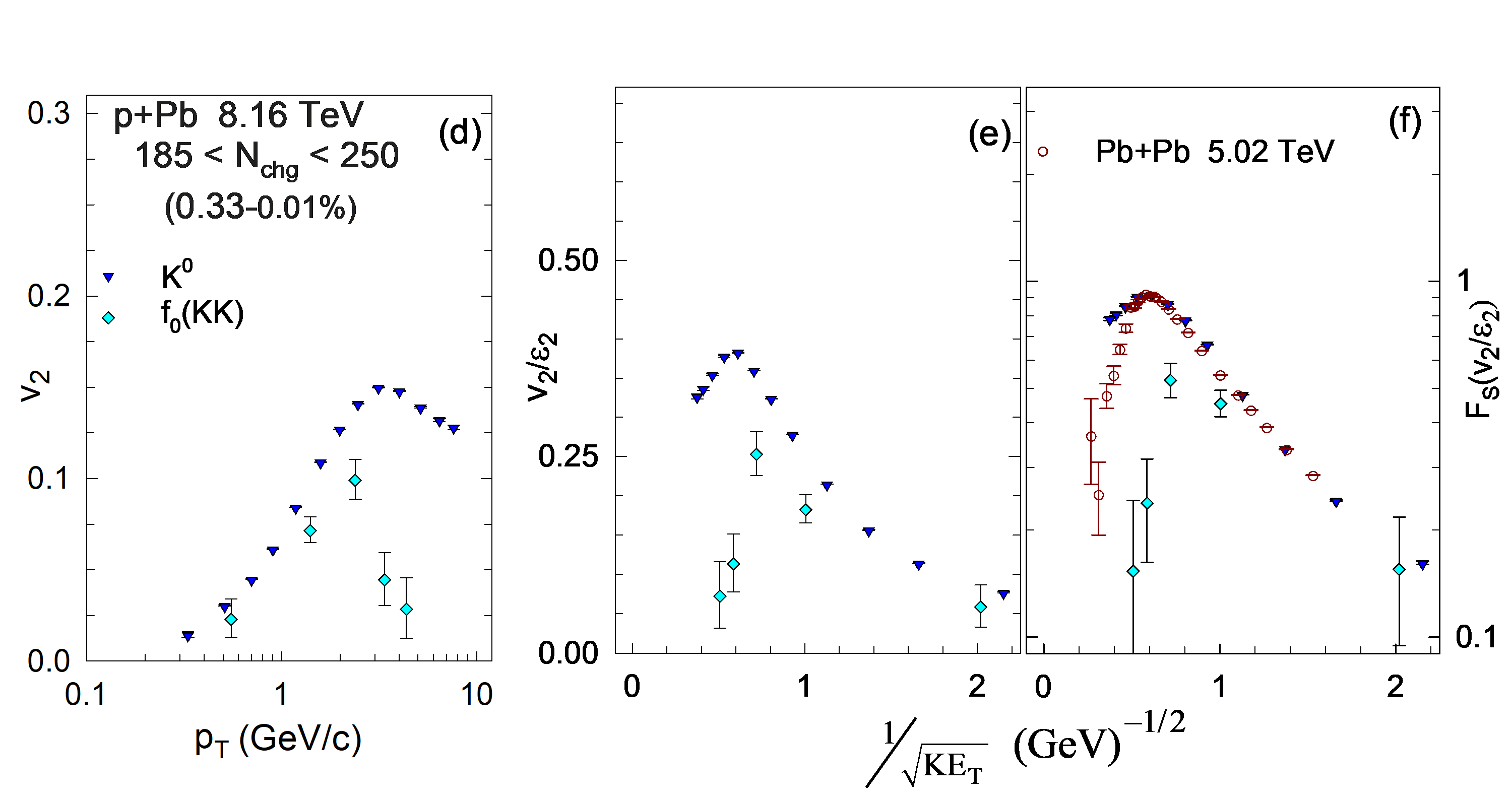}
\caption{Anisotropy-response scaling tests for the $f_0(980)$ in
high-multiplicity p+Pb collisions at $\sqrt{s_{NN}}=8.16$~TeV.
Panels (a)--(c) show the single-meson construction and panels
(d)--(f) the symmetric ${\rm K\bar K}$ constituent-response
construction. From left to right, the panels show $v_2(p_T)$,
$v_2/\varepsilon_2$, and the fully transformed responses compared
with the ultra-central Pb+Pb kaon reference.}
\label{fig:f0_structure_scaling}
\end{figure*}

\section{Results and Discussion}
\label{sec:results}
\subsection{Identified-hadron scaling in p+Pb}
\label{sec:pid_scaling_ppb}
Figure~\ref{fig:pid_scaling_ppb} illustrates the progressive scaling of
the identified-hadron $v_2$ measurements in high-multiplicity p+Pb
collisions at $\sqrt{s_{NN}}=8.16$~TeV. The raw $v_2(p_T)$ values in
Fig.~\ref{fig:pid_scaling_ppb}(a) exhibit pronounced species dependence.
Scaling by the initial eccentricity, shown in
Fig.~\ref{fig:pid_scaling_ppb}(b), reduces the response to
$v_2/\varepsilon_2$ but does not remove the separation among the
identified species.

Application of the full species-resolved response scaling produces the
collapse shown in Fig.~\ref{fig:pid_scaling_ppb}(c). The $K_S^0$,
$\Lambda$, $\Xi$, and $\Omega$ responses collapse onto the common
scaling function over a broad range of $1/\sqrt{{\rm KE}_T}$ after
the geometry, system-size, attenuation, radial-flow, and late-stage
hadronic contributions are scaled out. This closure establishes the
light-flavor system-response baseline used for the subsequent
$f_0(980)$ analysis.

The $D^0$ response provides a complementary heavy-flavor control.
Its scaling exhibits a distinct attenuation response relative to the
light-flavor mesons while preserving the same overall scaling
structure. It therefore provides a useful cross-check of the
species-resolved response framework without modifying the light-flavor
baseline used for the $f_0(980)$ analysis.

At larger ${\rm KE}_T$, departures from the common response become
increasingly apparent, particularly for the baryons, as the measurements
evolve away from the predominantly collective region toward a regime in
which the relative contributions from partonic energy loss and
jet-related correlations become increasingly important. The meson
response, however, remains substantially better scaled over the
corresponding momentum range than the baryon response, preserving the
relevant baseline for the subsequent $f_0(980)$ comparison. The
identified-hadron scaling therefore establishes the common p+Pb system
response together with the species-dependent departures from it at
larger ${\rm KE}_T$, providing the baseline against which the
$f_0(980)$ response is assessed.

\subsection{\texorpdfstring{$f_0(980)$}{f0(980)} response in p+Pb}
\label{sec:f0_scaling}
The single-meson and symmetric ${\rm K\bar K}$ constituent-response
constructions are compared directly in
Fig.~\ref{fig:f0_structure_scaling}. Panels (a)--(c) show the
single-meson construction, while panels (d)--(f) show the symmetric
${\rm K\bar K}$ construction. Both tests use the same measured
$f_0(980)$ points and the common system response established from the
identified-hadron scaling, which constrains the attenuation response
to $k_\beta\simeq1$. This common system response is held fixed when
the $f_0(980)$ is introduced.

Under the single-meson construction, the transformed $f_0(980)$
response in Fig.~\ref{fig:f0_structure_scaling}(c) exhibits good
scaling closure with the common response over the low-${\rm KE}_T$
region. At higher momentum, the $f_0(980)$ response follows the
corresponding evolution of the kaon response beyond the low-${\rm
KE}_T$ scaling region. Scaling closure is obtained with only a small
effective $f_0(980)$ final-state response,
$\zeta_{f_0}\simeq0.04$.

The symmetric ${\rm K\bar K}$ constituent-response construction gives
a distinctly different result. As shown in
Fig.~\ref{fig:f0_structure_scaling}(f), the transformed $f_0(980)$
response exhibits substantially poorer closure with the common kaon
response over the momentum region in which the established
light-flavor species scale. The measured $f_0(980)$ anisotropy
therefore follows the single-meson response substantially better than
the symmetric two-kaon constituent-response limit.

The response construction alone does not uniquely determine the
microscopic $f_0(980)$ structure. As demonstrated by the explicitly
molecular benchmark in Sec.~\ref{sec:molecular_benchmark},
single-meson-like scaling can also emerge from realistic
${\rm K\bar K}$ coalescence. The effective final-state response
therefore provides the complementary constraint.

For the CMS measurement, scaling closure under the single-meson
construction gives $\zeta_{f_0}\simeq0.04$, whereas the explicitly
molecular benchmark of Ref.~\cite{Wang:2025sxa} gives
$\zeta_{f_0}\simeq0.16$ after its common system response has been
independently constrained by the calculated kaons. For the CMS
measurement, the scaling fidelity is substantially better for
$\zeta_{f_0}\simeq0.04$ than for $\zeta_{f_0}\simeq0.16$,
demonstrating that the approximately fourfold difference is directly
resolved by the scaling. The measured $f_0(980)$ therefore exhibits
substantially weaker effective final-state sensitivity than the
molecular state represented by that calculation.

Taken together, the response construction and effective final-state
response provide a substantially more restrictive discriminator than
either scaling test alone. Within the framework tested here, the
measured $f_0(980)$ is characterized by a single-meson-like anisotropy
response with weak final-state sensitivity, distinctly different from
the explicitly molecular benchmark. This joint constraint provides a
new experimental discriminator of $f_0(980)$ formation dynamics and
places substantially stronger constraints on the long-standing
compact-versus-molecular ambiguity.

\subsection{Large-system scaling and implications for the
\texorpdfstring{$f_0(980)$}{f0(980)}}
\label{sec:large_system_implications}
\begin{figure*}[t]
\centering
\includegraphics[width=0.75\textwidth]
{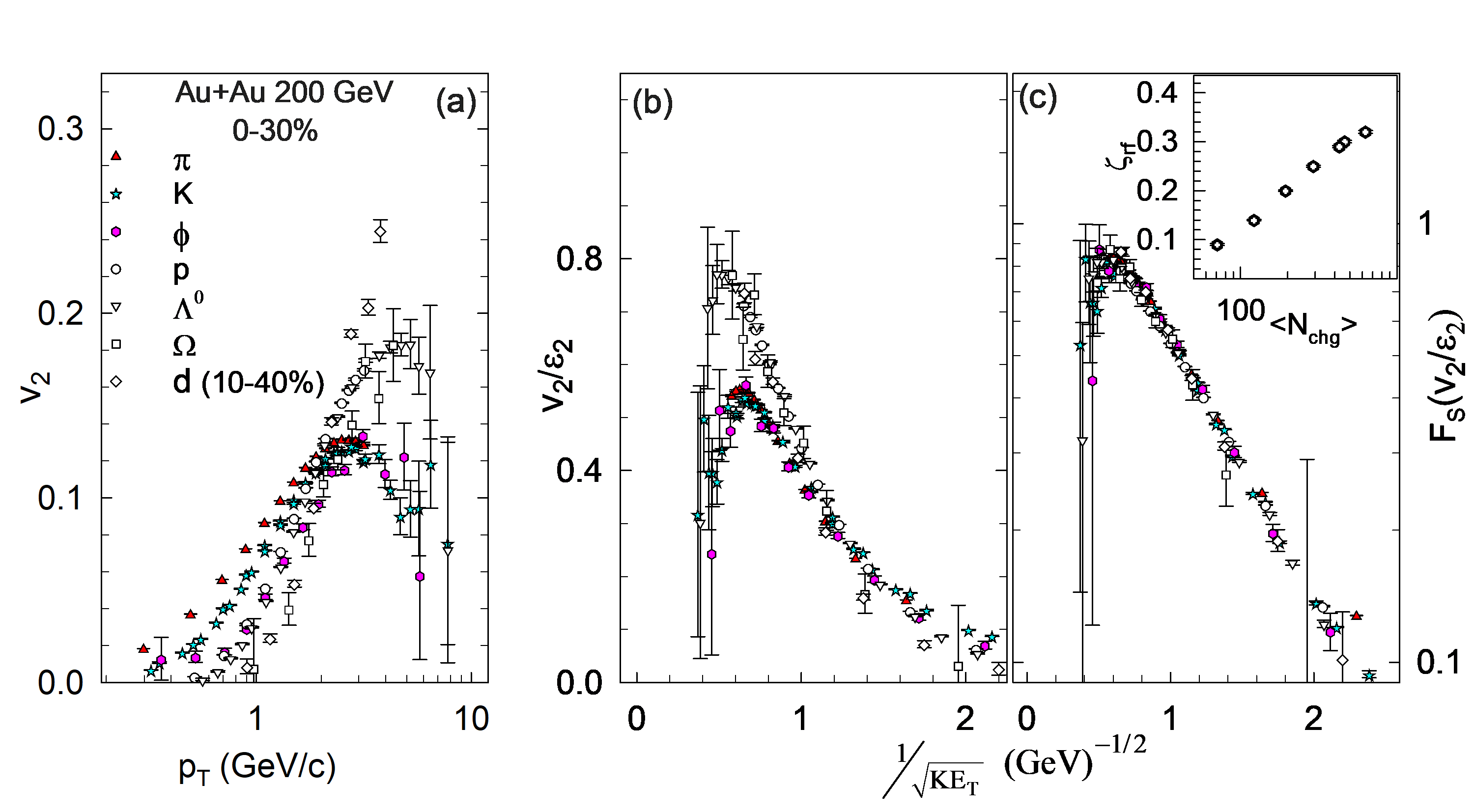}
\caption{Species-resolved anisotropy scaling for identified hadrons in
Au+Au collisions at $\sqrt{s_{NN}}=200$~GeV. The measurements are for
0--30\% central collisions, except for the deuteron, which is for
10--40\%. Shown are (a) $v_2(p_T)$, (b) the reduced responses
$v_2/\varepsilon_2$, and (c) the fully transformed responses; panels
(b) and (c) are shown as functions of $1/\sqrt{{\rm KE}_T}$.
The inset in panel (c) shows $\zeta_{\rm rf}$ as a function of
$\langle N_{\rm chg}\rangle$.}
\label{fig:pid_scaling_auau}
\end{figure*}

The species-resolved response scaling established in Pb+Pb collisions
at LHC energies persists in Au+Au collisions at
$\sqrt{s_{NN}}=200$~GeV. Figure~\ref{fig:pid_scaling_auau}, from
Ref.~\cite{Lacey:2024uky}, illustrates the progression of the scaling
for identified mesons and baryons at top RHIC energy. The raw
$v_2(p_T)$ measurements in Fig.~\ref{fig:pid_scaling_auau}(a) exhibit
pronounced species-dependent behavior. After scaling by the corresponding 
eccentricity and expressing the response as a function of $1/\sqrt{{\rm KE}_T}$,
Fig.~\ref{fig:pid_scaling_auau}(b) shows the identified species separating 
into approximate meson and baryon branches. The residual separation 
between these branches reflects, in particular, the species-dependent 
radial-flow response.

Application of the full meson and baryon response transformations
scales out this residual separation, bringing the identified species
onto the common response shown in
Fig.~\ref{fig:pid_scaling_auau}(c). The inset shows the corresponding
evolution of the radial-flow response parameter $\zeta_{\rm rf}$ with
$\langle N_{\rm chg}\rangle$. The scaling closure across identified
mesons and baryons demonstrates that the species-resolved framework
remains applicable from Pb+Pb collisions at LHC energies to Au+Au
collisions at top RHIC energy.

The established scaling in Pb+Pb and Au+Au provides a complementary
large-system baseline for extending the $f_0(980)$ response test
beyond p+Pb collisions. In particular, the $\phi$ meson provides a
stringent control because its mass is close to that of the $f_0(980)$.
Its successful mapping onto the common meson response substantially
reduces ambiguity associated with the leading kinematic mass
dependence and constrains the effective final-state response of a
meson at a comparable mass scale. An $f_0(980)$ measurement in a
large collision system can therefore be tested against a meson
response already established at a comparable mass scale.

The importance of such measurements extends beyond repeating the
p+Pb response-construction test. Pb+Pb and Au+Au collisions provide
complementary changes in both system size and beam energy relative to
the high-multiplicity p+Pb measurement. These changes can modify the
relative contributions from the partonic and hadronic stages, the
overall attenuation response, and the opportunities for late-stage
${\rm K\bar K}$ interaction, formation, and regeneration.

Within the present framework, these dependences can be tested through
the same complementary observables that distinguish the measured p+Pb
response from the explicitly molecular benchmark. Persistence of
single-meson scaling together with an effective final-state response
comparable to the p+Pb value would indicate that the weak final-state
sensitivity observed in p+Pb persists across these changes in the
collision environment. Conversely, an increase in $\zeta_{f_0}$ would
signal enhanced final-state sensitivity of the $f_0(980)$, while
evolution toward the symmetric ${\rm K\bar K}$
constituent-response limit would provide an independent indication of
increasing two-kaon response character.

Large-system $f_0(980)$ measurements therefore provide a particularly
incisive test of the interpretation obtained from p+Pb. The
species-resolved scaling already establishes the relevant meson and
baryon response in Pb+Pb and Au+Au before the $f_0(980)$ is
introduced, allowing a subsequent change in its response to be
distinguished from the underlying collision-system dependence. A
systematic comparison of the response construction and
$\zeta_{f_0}$ from p+Pb to Pb+Pb and Au+Au would consequently provide
complementary system-size and beam-energy tests of the $f_0(980)$
formation dynamics.

\section{Summary and Conclusions}
\label{sec:summary}
Species-resolved anisotropy-response scaling has been applied to the
$f_0(980)$ to investigate its structure and formation dynamics in
relativistic nuclear collisions. The analysis uses a common response
framework constrained by established identified mesons and baryons,
allowing the geometry, system size, attenuation, radial flow, and
hadronic re-scattering contributions to be scaled out before the
$f_0(980)$ response is examined. Two response constructions are
tested: a single-meson response and a symmetric ${\rm K\bar K}$
constituent-response limit. Their scaling fidelity, together with the
effective $f_0(980)$ final-state response $\zeta_{f_0}$ obtained from
scaling closure, provides complementary diagnostics of the
$f_0(980)$ formation dynamics.

In high-multiplicity p+Pb collisions at
$\sqrt{s_{NN}}=8.16$~TeV, the measured $f_0(980)$ exhibits broad
scaling closure under the single-meson construction, with
$\zeta_{f_0}\simeq0.04$ indicating only a small effective final-state
response. The symmetric ${\rm K\bar K}$ constituent-response
construction gives substantially poorer scaling closure. The measured
$f_0(980)$ anisotropy therefore follows the single-meson response
substantially better than the symmetric two-kaon constituent-response
limit.

The explicitly molecular ${\rm K\bar K}$ benchmark of
Ref.~\cite{Wang:2025sxa} demonstrates that single-meson-like scaling
alone is not a unique structural discriminator. Although the molecular
$f_0(980)$ exhibits broad scaling closure under the single-meson
construction, it gives a substantially larger effective final-state
response, $\zeta_{f_0}\simeq0.16$, after the common system response is
constrained by the calculated kaons. Its symmetric ${\rm K\bar K}$
construction shows approximate agreement at low momentum followed by
increasing departure from the common response. The response
construction and the magnitude of $\zeta_{f_0}$ must therefore be
considered jointly.

The combined constraints substantially reduce the ambiguity inherent
in either response construction alone. Within the framework tested
here, the CMS $f_0(980)$ is characterized by a single-meson-like
anisotropy response with weak final-state sensitivity, distinctly
different from the explicitly molecular benchmark.
Anisotropy-response scaling therefore provides a new experimental
discriminator of $f_0(980)$ formation dynamics based on the joint
constraints from the response construction and effective final-state
sensitivity rather than constituent counting alone.

The established species-resolved scaling in Pb+Pb and Au+Au provides
a controlled large-system baseline for extending this test beyond
p+Pb collisions. Measurements of the $f_0(980)$ anisotropy across
collision systems and beam energies would provide complementary
system-size and beam-energy lever arms for testing whether its response
construction and effective final-state response evolve with the
collision environment. Persistence of the p+Pb response pattern would 
indicate weak sensitivity to these changes. Conversely, an increase in
$\zeta_{f_0}$ would indicate enhanced final-state sensitivity, while
evolution toward the symmetric ${\rm K\bar K}$ constituent-response
limit would indicate increasing two-kaon response character and
sensitivity to late-stage two-kaon dynamics.

More broadly, these results establish anisotropy-response scaling as a
means of probing hadron formation dynamics beyond conventional
constituent-counting tests. By constraining the common collision-system
response with established identified particles before examining the
$f_0(980)$, the framework provides experimentally accessible
constraints on its formation dynamics and a systematic path for
testing their evolution across collision systems and beam energies.

%
%
\bibliography{f0_mech-refs}
\end{document}